\documentclass[aps,superscriptaddress,unsortedaddress,twocolumn,%
  showpacs,preprintnumbers,amsmath,amssymb]{revtex4-1}
\usepackage{graphicx}
\usepackage[dvipdfmx,breaklinks]{hyperref}

\newcommand{\bp}{\boldsymbol{p}}
\newcommand{\MeV}{\;\text{MeV}}
\newcommand{\GeV}{\;\text{GeV}}
\newcommand{\tr}{\text{tr}}

\begin{document}

\preprint{RBRC-951}
\title{Polyakov loop and QCD thermodynamics from the gluon and ghost
       propagators}

\author{Kenji Fukushima}
\affiliation{Department of Physics, Keio University, Kanagawa 223-8522,%
             Japan}

\author{Kouji Kashiwa}
\affiliation{RIKEN/BNL Research Center, Brookhaven National Laboratory,%
             Upton, NY 11973, USA}

\begin{abstract}
  We investigate quark deconfinement by calculating the effective
  potential of the Polyakov loop using the non-perturbative
  propagators in the Landau gauge measured in the finite-temperature
  lattice simulation.  With the leading term in the
  2-particle-irreducible formalism the resultant effective potential
  exhibits a first-order phase transitions for the pure SU(3)
  Yang-Mills theory at the critical temperature consistent with the
  empirical value.  We also estimate the thermodynamic quantities to
  confirm qualitative agreement with the lattice data near the
  critical temperature.  We then apply our effective potential to the
  chiral model study and calculate the order parameters and the
  thermodynamic quantities.  Unlike the case in the pure Yang-Mills
  theory the thermodynamic quantities are sensitive to the temperature
  dependence of the non-perturbative propagators, while the behavior
  of the order parameters is less sensitive, which implies the
  importance of the precise determination of the temperature-dependent
  propagators.
\end{abstract}

\maketitle


\paragraph*{Introduction}

It has been a long-standing question in physics of quantum
chromodynamics (QCD) how to understand confinement of quarks and
gluons in the vacuum and how to clarify the nature of deconfinement in
a medium in extreme environments (see
Ref.~\cite{Fukushima:2010bq,*Fukushima:2011jc} for reviews).

It was Polyakov~\cite{Polyakov:1978vu} who first addressed the
deconfinement phase transition successfully in the strong-coupling
limit of a pure Yang-Mills theory.  The order parameter for
deconfinement was then identified, which is now called the Polyakov
loop.  Later on, the strong-coupling expansion was extended to
implement quarks and the chiral
dynamics~\cite{Ilgenfritz:1984ff,*Gocksch:1984yk}.  One of the most
popular approaches in the QCD phase-diagram research, i.e.\ the chiral
effective model such as the Nambu--Jona-Lasinio
model~\cite{Fukushima:2004,Ratti:2006,Roessner:2006xn,Fu:2007xc,%
*Ciminale:2007sr,*Sakai:2008,*Fukushima:2008wg}, the linear-sigma
model~\cite{Schaefer:2007pw,*Schaefer:2009ui,*Herbst:2010rf,%
*Schaefer:2011ex}, etc~\cite{Megias:2006bn} with the Polyakov loop,
is a natural extension from the strong-coupling QCD.

The largest ambiguity in the P-chiral models lies in the choice of the
effective potential of the Polyakov loop.  Though the initial choice
was motivated by the strong-coupling expansion~\cite{Fukushima:2004},
it is now common to adopt the potential that fits the pure Yang-Mills
thermodynamics from the lattice simulation either in a polynomial
form~\cite{Ratti:2006} or in a Haar-measure
form~\cite{Fukushima:2004,Roessner:2006xn}.  Since the fitting
procedure does not refer to microscopic dynamics at all, it is unclear
how the Polyakov-loop potential is related or unrelated to
non-perturbative characteristics near $T_c$ as studied in the matrix
model~\cite{Dumitru:2010mj,*Dumitru:2012fw} (see also
Ref.~\cite{Sasaki:2012,*Ruggieri:2012} for an insight into the
physical contents of the potential).

An important breakthrough came from an attempt to understand quark
deconfinement in terms of the Landau-gauge propagators that describe
gluon confinement~\cite{Pawlowski:2003hq,Braun:2007bx}.  In the Landau
gauge it is the deep-infrared enhancement in the ghost propagator that
causes confinement, while the gluon propagator is infrared suppressed.
This behavior is qualitatively consistent with the confinement
scenarios by Kugo and Ojima (they are equivalent if the ghost
renormalization factor diverges at zero momentum~\cite{Kugo:1995km})
and also by Gribov and Zwanziger~\cite{Zwanziger:1993dh}.  Indeed the
gluon and ghost propagators in the Landau gauge at zero and finite
temperature have been calculated in the lattice
simulation~\cite{Bonnet:2000,*Gattnar:2004bf,*Cucchieri:2007b,%
*Bogolubsky:2007,*Iritani:2009mp,Cucchieri:2007a,%
*Sternbeck:2007,Aouane:2011fv}, 
the Dyson-Schwinger equation
(DSE)~\cite{Smekal:1997is,*Smekal:1997vx,*Alkofer:2000wg,*Fischer:2002a,%
*Fischer:2002b,*Fischer:2006vf,*Dudal:2007,*Aguilar:2008,*Boucaud:2008,%
*Dudal:2008,*Binosi:2009,*Boucaud:2011,Fischer:2008uz},
the functional renormalization group (FRG)
approach~\cite{Braun:2007bx,Fischer:2008uz,Marhauser:2008fz,*Braun:2009gm,%
*Kondo:2010ts}.

In this work we report on an extension of Ref.~\cite{Braun:2007bx}
using the $T$-dependent propagators from the lattice-QCD
simulation~\cite{Aouane:2011fv} (see
Ref.~\cite{Fister:2011um,*Fister:2011uw} for the finite-$T$
propagators from the FRG calculation).  Also, we try to address the
thermodynamic quantities constructed from the propagators.  We would
stress here that our central point is not only an extension of
Ref.~\cite{Braun:2007bx} but it should aim to build a bridge over the
model studies and the first-principle functional approaches.  Such a
work must be extremely useful for both sides; there are many arguments
to suggest that the back-reaction from the matter to the glue sector
has crucial impacts on the QCD phase-diagram research based on the
P-chiral models~\cite{Schaefer:2007pw,Fukushima:2010is}.  In principle
this shortcoming of insufficient back-reaction in the P-chiral models
could be resolved with the DSE or FRG
approaches~\cite{Pawlowski:2010ht}, but then one has to confront
huge-scale computations.  As we see later, thanks to a nice
parametrization of the propagators in Ref.~\cite{Aouane:2011fv}, our
practical strategy requires only minimal modifications in the
model-study procedures.  The outcomes are quite promising as we
will see.
\\


\paragraph*{Propagator and Parametrization}

In the covariant gauge the gluon propagator inverse can be decomposed
into the transverse and the longitudinal parts using the projection
operators;
\begin{equation}
 T_{\mu\nu} \equiv g_{\mu\nu}-\frac{p_\mu p_\nu}{p^2}\;,\qquad
 L_{\mu\nu} \equiv g_{\mu\nu}-T_{\mu\nu}\;,
\end{equation}
and the further decomposition relative to the rest frame at finite
temperature,
\begin{equation}
 P^T_{\mu\nu} \equiv (1\!-\!\delta_{\mu 4})(1\!-\!\delta_{\nu 4})
  \Bigl( \delta_{\mu\nu}\!-\!\frac{p_\mu p_\nu}{\bp^2} \Bigr),\quad
 P^L_{\mu\nu} \equiv T_{\mu\nu}\!-\! P^T_{\mu\nu}.
\end{equation}
The gluon propagator in the Landau gauge is then parametrized as
\begin{equation}
 D_{A\,\mu\nu}(p^2) = \delta^{ab} \Bigl( D_T^{(T)}P^T_{\mu\nu}
  +D_T^{(L)}P^L_{\mu\nu} + \xi D_LL_{\mu\nu} \Bigr)\;.
\end{equation}
Interestingly these propagators as well as the ghost propagator $D_C$
are compactly expressed in a Gribov-Stingl form;
\begin{equation}
 \begin{split}
 & D_T^{(T)} \propto \frac{d_t (p^2+d_t^{-1})}{(p^2+r_t^2)^2}\;,\qquad
  D_L = \frac{1}{p^2}\;,\\
 & D_T^{(L)} \propto \frac{d_l (p^2+d_l^{-1})}{(p^2+r_l^2)^2}\;, \qquad
  D_C = \frac{p^2 + d_g^{-1}}{(p^2)^2}\;,
 \end{split}
\label{eq:fit}
\end{equation}
\begin{table}
\begin{tabular}{c|cccc}
 $T$ & ~$r_t^2\;({\rm GeV}^2)$~ & ~$d_t\;({\rm GeV}^{-2})$~
  & ~$r_l^2\;({\rm GeV}^2)$~ & ~$d_l\;({\rm GeV}^{-2})$~ \\ \hline
 $0.86T_c$ & 0.880 & 0.143 & 0.256 & 0.220 \\
 $1.20T_c$ & 0.963 & 0.140 & 1.018 & 0.162
\end{tabular}
\caption{Parameters for the non-perturbative propagators at
  $T=0.86T_c$ and $1.20T_c$ taken from Ref.~\cite{Aouane:2011fv}.}
\label{tab:parameter}
\end{table}
as discussed in
Refs.~\cite{Cucchieri:2003,*Cucchieri:2011di,Aouane:2011fv}.  
This form also appears in a low-energy effective description of the
Yang-Mills theory~\cite{Kondo:2011ab}.  We note that we postulated
non-renormalization for the longitudinal gluon propagator because
$Z_L = 1 + \mathcal{O}(\xi)$ and $\xi\to 0$ is taken after all in the
Landau gauge.  In this work we use two sets of parameters at
$T=0.86 T_c$ and $T=1.20 T_c$ taken from Ref.~\cite{Aouane:2011fv},
which is listed in Tab.~\ref{tab:parameter}.  The ghost propagator can
be well reproduced by a choice of $d_g^{-1}=0.454\GeV^2$ which is
nearly independent of $T$.
\\


\paragraph*{Polyakov Loop Potential}

The Gribov-Stingl form is especially convenient for the practical
computation of the partition function and the effective potential of
the Polyakov loop.  In the present study, we make use of an
approximation motivated from the 2-particle-irreducible (2PI)
formalism or the FRG equation.  In the 2PI formalism for example, the
full effective action can be expressed generally as
\begin{equation}
 \Gamma = \frac{1}{2}\tr\ln G^{-1}
  -\frac{1}{2}\tr\ln(G^{-1}-G_0^{-1})G +\Gamma_2[G]
\label{eq:2PI}
\end{equation}
with the full propagator $G$, the tree-level propagator $G_0$, and the
sum of the 2PI diagrams $\Gamma_2[G]$.  Thus, all known approximations
are derived from some truncation onto Eq.~\eqref{eq:2PI};  the Hartree
approximation for example picks up the first term $\tr\ln G^{-1}$ only
with an assumption that $G$ is written with a mean-field mass.
Usually the full propagator should be fixed self-consistently within
the 2PI formalism.  Now that we already know the non-perturbative
propagators, however, we can directly plug them into
Eq.~\eqref{eq:2PI} to have better results.  In this ``hybrid'' method,
the dominant contribution to the gluonic pressure comes from
\begin{align}
 & \beta \Omega_{\text{glue}} \simeq -\frac{1}{2}\tr\ln D_A^{-1}
  + \tr\ln D_C^{-1} \notag\\
 &= -\frac{1}{2}\tr\ln D_L^{-1}  \!-\! \tr\ln D_T^{(T)-1}
    \!-\! \frac{1}{2}\tr\ln D_T^{(L)-1} \!+\! \tr\ln D_C^{-1} \;.
\label{eq:approx}
\end{align}
We note that this leading-order truncation corresponds to the
so-called quasi-particle model of
thermodynamics~\cite{Zwanziger:2004np}.  In other words this method
presumably works in the deconfinement regime where gluons are the
physical degrees of freedom, while the approximation may fail the
regime of glueballs at low $T$.  Importantly, in principle, one can
systematically improve the approximation by evaluating the sub-leading
terms and $\Gamma_2$ diagrammatically once $D_A$ and $D_C$ are known.

For the purpose of calculating the effective potential for the
Polyakov loop, we keep a constant $A_4$ in the temporal covariant
derivative;  $p^2$ in the gluon and ghost propagators are replaced
with $\tilde{p}^2 = (2\pi Tn + g\beta A_4)^2+\bp^2$ in the color
adjoint representation.  We note that the Polyakov loop is defined as
\begin{equation}
 \Phi \equiv \frac{1}{3}\tr L_3 = \frac{1}{3}\tr\,
  \mathcal{P} \exp\biggl(ig\int_0^\beta A_4\,dx_4\biggr) \;,
\end{equation}
where $\mathcal{P}$ represents the time ordering and $A_4$ is a matrix
in the color fundamental representation.

Then, because the propagators~\eqref{eq:fit} are written as a
combination of $(p^2+m^2)^{-1}$, it is straightforward to carry out
the summation over the Matsubara frequency in an analytical way,
i.e.\ for the transverse gluons for example, we have
\begin{align}
 \tr\ln D_T^{(T)-1} &= 2 \tr\ln(\tilde{p}^2+r_t^2)
  - \tr\ln(\tilde{p}^2+d_t^{-1}) \notag\\
  &= 2W_B(r_t^2) - W_B(d_t^{-1}) \;,
\label{eq:DT}
\end{align}
where we have defined,
\begin{equation}
 W_B(m^2) \equiv -2V\int\frac{d^3 p}{(2\pi)^3}\, \tr\ln\Bigl(1-
  L_8 \,e^{-\beta\sqrt{\bp^2+m^2}}\Bigr)
\label{eq:WB}
\end{equation}
with $L_8$ being the Polyakov loop matrix in the color adjoint
representation; $(L_8)_{ab}=2\tr(t_a L_3 t_b L_3^\dagger)$.  We
dropped the divergent zero-point energy that is independent of $T$ and
thus an irrelevant offset.  One can take the trace in color space
explicitly to
find~\cite{Sasaki:2012},
\begin{equation}
 W_B(m^2) = -2 V \int\frac{d^3 p}{(2\pi)^3}
  \ln\biggl(1+\sum_{n=1}^8 C_n e^{- n \beta\sqrt{\bp^2+m^2}}\biggr) \;,
\label{eq:WB2}
\end{equation}
where $C_8 = 1,\; C_1 = C_7 = 1 - 9 {\bar \Phi} \Phi,\;
C_2 = C_6 = 1 - 27 {\bar \Phi} \Phi + 27 ( {\bar \Phi}^3 + \Phi^3 ),\;
C_3 = C_5 = -2 + 27 {\bar \Phi} \Phi - 81 ({\bar \Phi} \Phi)^2,\;
C_4 = 2 \Bigl[ - 1 + 9 {\bar \Phi}\Phi -27 ({\bar \Phi}^3 + \Phi^3) 
                + 81 ({\bar \Phi}\Phi)^2 \Bigr]$.

\begin{figure}
 \includegraphics[width=0.85\columnwidth]{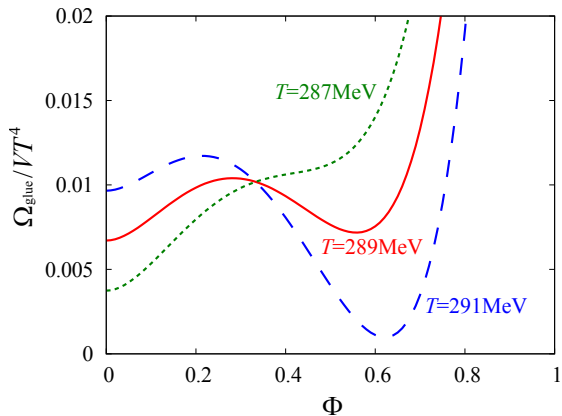}
 \caption{Polyakov loop potential near $T_c$ that exhibits a
   first-order phase transition as expected for the SU(3) case.  With
   the parameter sets at $T=0.86T_c$ and $T=1.20T_c$ in
   Tab.~\ref{tab:parameter} we find $T_c=289\MeV$ (as shown) and
   $T_c=351\MeV$, respectively.}
 \label{fig:poten}
\end{figure}

In this way, using Eqs.~\eqref{eq:approx}, \eqref{eq:DT}, and
\eqref{eq:WB2}, we can numerically calculate the Polyakov-loop
effective potential $\beta\Omega_{\text{glue}}[\Phi]$, and then
determine $\Phi$ from
$\partial\Omega_{\text{glue}}[\Phi]/\partial\Phi=0$.  We find that the
critical temperature in our treatment is $T_c=289\MeV$ and
$T_c=351\MeV$ for the parameter sets at $T=0.86T_c$ and $T=1.20T_c$ in
Tab.~\ref{tab:parameter}, respectively.  If we utilize the full
$T$-dependent propagator, thus, the critical temperature should be
somewhere in between.  Considering the empirical value
$T_c\sim 280\MeV$~\cite{Boyd:1996,*Kaczmarek:2002}, one might wonder
that our estimate of $T_c$ is a bit too larger.  This is fine because
neglected $\Gamma_2$ has an effect to push $T_c$
down~\footnote{Private communications with Jan~Pawlowski.}.  This
semi-quantitative agreement of $T_c$ is amazing for its simplicity.
The energy scales are provided through $r^2$ and $d$ as given in
Tab.~\ref{tab:parameter} and there is no adjustable degrees of
freedom.  It is also worth mentioning that the Polyakov loop potential
formulated here leads to a second-order phase transition for the color
SU(2) case.

Although $T_c$ is such different depending on the parameter sets,
interestingly, $\Phi$ and all thermodynamic quantities as functions of
$T$ are \textit{identical} if plotted in the unit of $T_c$.  This is a
quite non-trivial finding, and supports the validity of the fitted
Polyakov loop potential characterized by only one dimensionful
parameter $T_0$~\cite{Ratti:2006,Roessner:2006xn}.

\begin{figure}
 \includegraphics[width=0.9\columnwidth]{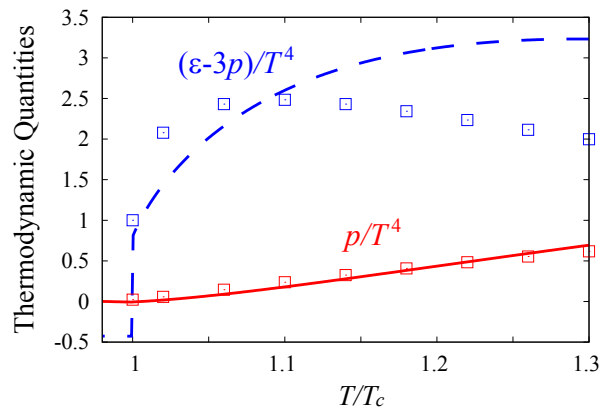}
 \caption{Thermodynamic quantities as compared to the lattice results
   taken from a table in Ref.~\cite{Borsanyi:2012ve}.  The numerical
   results are indistinguishable for the parameter sets of $T=0.86T_c$
   and $T=1.2T_c$.}
 \label{fig:thermo}
\end{figure}

We show thermodynamic quantities as a function of $T$ in
Fig.~\ref{fig:thermo} where the normalized pressure and the
interaction measure are plotted.  Because we adopt the parameter sets
at $T=0.86T_c$ and $T=1.20T_c$ (yielding indistinguishable results if
scaled with $T_c$), we choose the temperature range up to $1.3T_c$
here.  We make a comparison with the lattice data and confirm
semi-quantitative agreement.  In particular, as seen in
Fig.~\ref{fig:thermo}, the agreement of the pressure is pretty good,
while the interaction measure does not fit the lattice data well.

The discrepancy above $\sim 1.2T_c$ should be attributed to the
neglected $T$-dependence in the non-perturbative propagators as seen
in Refs.~\cite{Aouane:2011fv,Fischer:2010}.  Once it is correctly
taken into account, we expect that the overshoot of thermodynamic
quantities could become milder.  It should be mentioned that, in the
$T\to\infty$ limit, thermodynamic quantities in our method approach
the Stefan-Boltzmann limit, by construction, as all propagators go to
$\sim 1/p^2$.  Below $T_c$, on the other hand, we find that some of
thermodynamic quantities go negative, which is caused by too strong
ghost contributions.  It is an unanswered question how to extract the
expected behavior of the glueball
gas~\cite{Borsanyi:2012ve,Sasaki:2012} from the gluon and the ghost
propagators.  One should cope with $\Gamma_2$ in Eq.~\eqref{eq:2PI}
in the confinement regime.  Although one may think that glueballs are
too heavy to make a sizable contribution to thermodynamics, the
electric glueballs can be significantly light in the vicinity of the
(second-order) critical point~\cite{Ishii:2001zq,*Hatta:2003ga}.
\\


\paragraph*{Dynamical Quarks}

Just for the demonstration purpose of the usefulness of our effective
potential let us apply it to an effective model.  We emphasize that
our goal is not to analyze the model itself but to seek for the
possibility to utilize this \textit{inverted Weiss potential} in
phenomenology that is complementary to the field-theoretical
argument~\cite{Braun:2011fw,*Braun:2012zq}.  To this end we adopt the
covariant coupling in the (2+1)-flavor quasi-quark description.  Then,
the thermodynamic potential from the quark contribution reads,
\begin{align}
  & \beta\Omega_{\text{quark}} = -6 \beta V\sum_f \int^\Lambda
   \frac{d^3p}{(2\pi)^3} \sqrt{\bp^2+M_f^2}
   - 4\sum_f W_F(M_f^2) \notag\\
  & \quad
   + g_s\bigl(\langle\bar{u}u\rangle^2 + \langle\bar{d}d\rangle^2
   + \langle\bar{s}s\rangle^2 \bigr)
   + 4g_d \langle\bar{u}u\rangle \langle\bar{d}d\rangle
     \langle\bar{s}s\rangle\;,
\label{eq:wf}
\end{align}
where $M_u=m_u-2g_s\langle\bar{u}u\rangle-2g_d\langle\bar{d}d\rangle
\langle\bar{d}d\rangle$ etc and the last two terms above represent the
condensation energy.  Here we defined,
\begin{equation}
 W_F(M^2) \equiv V\int\frac{d^3 p}{(2\pi)^3}\, \tr\ln\Bigl( 1
  + L_3\, e^{-\beta\sqrt{\bp^2+M^2}} \Bigr) \;,
\end{equation}
in the color fundamental representation.  Thus, we choose the
parameters according to the standard set of the (2+1)-flavor NJL model
as $\Lambda=631.5\MeV$, $g_s\Lambda^2=3.67$,
$g_d\Lambda^5=-9.29$, $m_u=m_d=5.5\MeV$,
$m_s=135.7\MeV$~\cite{Hatsuda:1994pi}.

\begin{figure}
 \includegraphics[width=0.9\columnwidth]{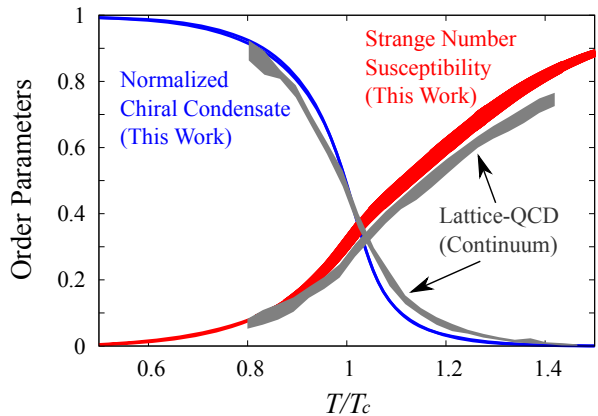}
 \caption{Order parameters (see Eqs.~\eqref{eq:delta} and
   \eqref{eq:chi_s} for definition) from our model calculation and the
   lattice-QCD results.  The bands in our results correspond to the
   uncertainty from the parameter choices at $T=0.86T_c$ (with
   $T_c=182.5\MeV$ for the unit of the horizontal axis) and
   $T=1.20T_c$ (with $T_c=191\MeV$).  The gray bands are the
   lattice-QCD results in the continuum extrapolation from
   Refs.~\cite{Borsanyi:2010a} (with $T_c=156\MeV$).}
 \label{fig:orderfull}
\end{figure}

We can express $W_F(M^2)$ in terms of $\Phi$~\cite{Ratti:2006} and
solve the Polyakov loop $\Phi$ and the chiral condensates
$\langle\bar{q}q\rangle$ to minimize the total potential
$\Omega_{\text{glue}}+\Omega_{\text{quark}}$.
Figure~\ref{fig:orderfull} shows our numerical results for the order
parameters;  The normalized chiral condensate is defined as
\begin{equation}
 \Delta = \frac{\langle\bar{u}u\rangle
  - (m_{ud}/m_s)\langle\bar{s}s\rangle}{\langle\bar{u}u\rangle_0
  - (m_{ud}/m_s)\langle\bar{s}s\rangle_0} \;,
\label{eq:delta}
\end{equation}
where $\langle\bar{u}u\rangle=\langle\bar{d}d\rangle$ represents the
light-quark chiral condensate and the denominator is the value at
$T=0$.  We adopt the strange-quark number susceptibility,
\begin{equation}
 \chi_s/T^2 = \frac{1}{T^2}\frac{\partial n_s}{\partial \mu_s}
  = -\frac{1}{VT^2}\frac{\partial^2 \Omega}{\partial \mu_s^2}
\label{eq:chi_s}
\end{equation}
for the deconfinement order parameter instead of the conventional
Polyakov loop.  This is because the Polyakov loop has a large
renormalization factor, and in the mean-field approximation $\Phi$
should be considered as an internal parameter rather than an
observable.  To draw Fig.~\ref{fig:orderfull} we chose $T_c=182.5\MeV$
for the parameter set at $T=0.86T_c$, $T_c=191\MeV$ for that at
$T=1.20T_c$, and the lattice-QCD results from
Ref.~\cite{Borsanyi:2010a} are plotted as a function of $T/T_c$ with
$T_c=156\MeV$.  One might have an impression that $T_c$ in the model
side is too large as compared to the lattice-QCD data, but this is to
be improved with the back-reaction from the polarization
diagrams~\cite{Schaefer:2007pw,Fukushima:2010is}.

\begin{figure}
 \includegraphics[width=0.9\columnwidth]{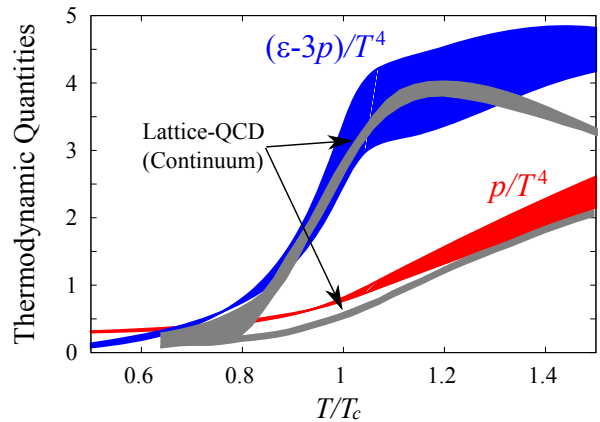}
 \caption{Thermodynamic quantities as compared to the lattice-QCD
   results taken from Ref.~\cite{Borsanyi:2010b}.  The bands
   correspond to the uncertainty from the parameter choices at
   $T=0.86T_c$ and $T=1.20T_c$.  The choice of $T_c$ is the same as
   in Fig.~\ref{fig:orderfull}.}
 \label{fig:thermofull}
\end{figure}

In this case with dynamical quarks thermodynamic quantities do not
show unphysical behavior near $T_c$ because quark degrees of freedom
dominate over gluons.  The pressure and the interaction measure are
plotted in Fig.~\ref{fig:thermofull}.  We see that our numerical
results quantitatively agree with the lattice-QCD data taken from
Ref.~\cite{Borsanyi:2010b}.

Our method with the momentum integration in Eq.~\eqref{eq:WB} is
simple enough to be an alternative of the Polyakov-loop potential used
in the market of the P-chiral models.  Moreover, it is advantageous
for our method to be extendable to implement missed contributions such
as the screening effects through the quark polarization diagrams.
The improvement in this direction should be extremely important to
tackle the realistic QCD phase structure especially at high baryon
density or strong magnetic field~\cite{Kashiwa:2011js} (for a recent
review, see Ref.~\cite{Gatto:2012sp}).  These effects do not directly
couple to gluons, and nevertheless, the nature of deconfinement is
affected substantially through the quark loops that carry the baryon
number and the electric charge.  Progresses in this direction shall be
reported elsewhere.
\\


\paragraph*{Summary}

We elucidated how to construct the effective potential of the Polyakov
loop from the non-perturbative propagators of gluon and ghost in the
Landau gauge available from the lattice simulation.  This is an
extension of the idea of Ref.~\cite{Braun:2007bx}.  We took the
fitting forms of the finite-temperature propagators from
Ref.~\cite{Aouane:2011fv} and found a quite tractable way to calculate
thermodynamic quantities as well as the order parameters as functions
of $T$.  We showed that the thermodynamic properties are nicely
consistent with the lattice data in the vicinity of $T_c$.
Furthermore we introduced dynamical quarks in the quasi-quark
approximation to reproduce the simultaneous crossovers of
deconfinement and chiral restoration.  We made sure that our potential
works well even on the quantitative level without fine-tuning of any
parameter.  It would be an intriguing future problem to apply our
Polyakov-loop potential to the non-local version of the chiral
model~\cite{Kashiwa:2011}.

We believe that this present work takes one step forward to the
understanding of the QCD phase diagram in extreme environments based
on the first-principle-type calculations.

\begin{acknowledgments}
The authors thank Wolfram~Weise for kind hospitality at TUM where this
work was initiated.  They also thank Jens~Braun, David~Dudal,
Michael~Ilgenfritz, and Marco~Ruggieri for comments.  K.F.\ thanks
Jan~Pawlowski and Nan~Su for useful discussions.  K.K.\ is supported
by RIKEN Special Postdoctoral Researchers Program.  K.F.\ is supported
by Grant-in-Aid for Young Scientists B (24740169).
\end{acknowledgments}

\bibliography{Note-TC}

\end{document}